\documentclass [12pt] {IEEEoj}
\usepackage{fix-cm}
\usepackage{cite}
\usepackage{balance}
\usepackage{amsmath,amssymb,amsfonts}
\usepackage{algorithmic}
\usepackage{graphicx,color}
\usepackage{textcomp}
\usepackage{booktabs}
\def\BibTeX{{\rm B\kern-.05em{\sc i\kern-.025em b}\kern-.08em
    T\kern-.1667em\lower.7ex\hbox{E}\kern-.125emX}}
\AtBeginDocument{\definecolor{ojcolor}{cmyk}{0.93,0.59,0.15,0.02}}
\receiveddate{XX Month, XXXX}
\reviseddate{XX Month, XXXX}
\accepteddate{XX Month, XXXX}
\publisheddate{XX Month, XXXX}
\currentdate{XX Month, XXXX}
\doiinfo{}

\begin{document}
\bstctlcite{IEEEexample:BSTcontrol}

\title{A Novel Low-Complexity Peak-Power-Assisted Data-Aided Channel Estimation Scheme for MIMO-OFDM Wireless Systems}

\author{INAAMULLAH KHAN, MOHAMMAD MAHMUDUL HASAN, AND MICHAEL CHEFFENA}
\affil{Faculty of Engineering, Norwegian University of Science and Technology (NTNU), 2815 Gjøvik, Norway}

\corresp{CORRESPONDING AUTHOR: I. KHAN (e-mail: inaamullah.khan@ntnu.no)}
\authornote{This work was supported by the Norwegian University of Science and Technology.}

\begin{abstract}
This paper, for the first time, presents a low-complexity peak-power-assisted data-aided channel estimation (DACE) scheme for both single-input single-output (SISO) and multiple-input multiple-output orthogonal frequency division multiplexing (MIMO-OFDM) wireless systems. In OFDM, high peak-power levels occur when the subcarriers align in phase and constructively interfere with each other. The research proposes a peak-power-assisted channel estimation scheme that accurately selects peak-power carriers at the transmitter of an OFDM system and uses them as reliable carriers for the DACE scheme. By incorporating these reliable carriers with known pilot symbols as additional pilot signals, channel estimation accuracy significantly improves in MIMO-OFDM systems. This eliminates the need to determine reliable data symbols at the receiver, thereby significantly reducing the computational complexity of the system. However, high peak-powers are considered a major drawback in OFDM. In this work, we incorporate a companding technique to mitigate this issue and provide sufficient margin for the DACE scheme. The performance of the proposed DACE scheme is evaluated using both least square (LS) and linear minimum mean square error (LMMSE) channel estimators. In this regard, the proposed technique not only improves channel estimation accuracy but also enhances the spectral efficiency of the wireless system. It outperforms traditional channel estimators in terms of system mean square error (MSE) and bit-error-rate (BER) performance. It also reduces the pilot overhead by 50$\%$ compared to traditional channel estimators and provides bandwidth optimization for MIMO-OFDM systems. This makes it a promising solution for improving the performance and efficiency of wireless communication systems.
\end{abstract}


\maketitle

\section{INTRODUCTION}
\subsection{BACKGROUND AND MOTIVATION}
\IEEEPARstart{O}{ver} the years, there have been significant improvements in wireless communication systems \cite{ref1}, \cite{ref2}. To meet the increasing demands for high data rates, greater reliability, and enhanced capacity, orthogonal frequency division multiplexing (OFDM) has emerged as a key technology in 4G and 5G wireless communication systems. It is a multicarrier modulation (MCM) technique that divides the available spectrum into a number of overlapping but orthogonal subchannels. This provides resilience to multipath fading and enables higher spectral efficiency in wireless communication systems \cite{ref3},\cite{ref4}. However, ensuring accurate channel state information (CSI) is crucial for coherent signal detection in OFDM-based systems \cite{ref5}. Therefore, the quality of channel estimation becomes critical to achieve improved system performance. However, obtaining perfect CSI is highly complex due to the rapid variations in the wireless channel \cite{ref6}.

Channel estimation is one of the most challenging issues in multiple-input multiple-output orthogonal frequency division multiplexing (MIMO-OFDM) wireless systems \cite{ref7}, \cite{ref8}. Several research efforts have been made and different channel estimation schemes have been proposed to achieve better system performance \cite{ref9}. However, least square (LS) and linear minimum mean square error (LMMSE) methods are used and optimized under different conditions, in particular within pilot-assisted channel estimation (PACE) schemes. The LS estimator requires no a priori knowledge of the channel statistics, making it computationally less complex and efficient at high signal-to-noise ratio (SNR). The LMMSE, on the other hand, requires second-order moments for both channel and noise to minimize the mean square error (MSE). It is therefore offers better system performance at the cost of higher computational complexity \cite{ref10}, \cite{ref11}.

Regardless of the channel estimation method used, a key issue is how to improve  channel estimation accuracy without increasing the number of pilots in MIMO-OFDM systems. Numerous studies have addressed this issue and have shown that data-aided channel estimation (DACE) techniques generally result in improved system performance. These techniques select the data carriers least affected by the wireless channel and utilize them as additional pilot signals. By incorporating these reliable data carriers with known pilot symbols, the spectral efficiency of MIMO-OFDM systems is enhanced \cite{ref12}. However, another challenge is to develop an algorithm that can intelligently identify the reliable data carriers. In this regard, only a few detection methods have been discussed in the literature, such as maximum likelihood (ML), which is an optimal estimator. However, such a hard-decoding scheme is efficient at high SNR in the classical additive white Gaussian noise (AWGN) scenario \cite{ref13}. On the other hand, the maximum a-posteriori probability (MAP) data detector is used for soft symbol detection. It computes a-posteriori probabilities (APPs) for the given received signals, and a threshold for APPs is set to select the detected symbols that are less affected by the wireless channel. However, such algorithms have higher computational complexity.

In general, a large number of closely spaced pilot symbols are required to capture fast variations in the wireless channel, but this reduces the spectral efficiency of the wireless systems. On the other hand, a small number of pilots cannot accurately sample the fading process, thus deteriorate the quality of channel estimation. To address this trade-off, researchers have proposed various iterative and non-iterative DACE schemes to improve channel estimation quality without compromising the spectral efficiency of the wireless systems. However, non-iterative DACE schemes suffer from error propagation due to data detection errors. In contrast, iterative DACE schemes are computationally complex and can cause communication latency \cite{ref14}. Therefore, we proposed a novel DACE scheme for both single-input single-output (SISO) and MIMO-OFDM systems to accurately detect reliable data symbols out of all possible symbols transmitted through the wireless channel \cite{ref15}. The proposed algorithm considered both noise and channel estimation error to eliminate the possibility of error in the detection process. However, the receiver-based DACE scheme requires higher computational complexity as it calculates the reliability of each observation relative to the nearest constellation point as well as all the nearest neighbours. It therefore, offers improved system performance at a significantly higher computational cost. Motivated by this, in this work, we propose a novel low-complexity peak-power-assisted DACE scheme for both SISO and MIMO-OFDM systems.

\subsection{RELATED WORKS}
DACE schemes have been developed to overcome the limitations of PACE schemes. PACE schemes rely solely on pilot symbols, which reduces the spectral efficiency of the wireless systems. In contrast, DACE schemes utilize both pilots and decoded data symbols to improve channel estimation accuracy while maintaining higher spectral efficiency. They are more robust in dynamic environments, as they adaptively refine channel estimates by leveraging information from decoded data. This leads to better MSE and bit-error-rate (BER) performance, particularly in scenarios with high mobility or severe fading. Furthermore, DACE schemes reduce the dependency on the density and placement of pilot symbols, enabling more flexible and efficient resource allocation. Consequently, researchers have proposed various DACE schemes to accurately estimate the wireless channel.

In \cite{ref14}, a data-aided LMMSE channel estimator is proposed for MIMO systems, which selectively exploits detected symbol vectors as additional pilot signals. In this work, a Markov decision process (MDP) is defined to optimize the selection of detected symbol vectors, aiming to minimize the MSE of the channel estimate. Simulation results demonstrate that the detection performance is improved compared to that of conventional LMMSE channel estimators. In \cite{ref15}, a novel DACE algorithm is proposed for both SISO and MIMO-OFDM systems. The proposed estimator selects the most reliable data carriers with high accuracy and significantly reduces the pilot overhead to improve the spectral efficiency of MIMO-OFDM systems. Similarly, in \cite{ref16}, another DACE scheme is introduced that leverages both clustering and reinforcement learning techniques for MIMO systems. The proposed scheme selectively utilizes detected symbols that are less affected by destructive interference as additional pilot signals. However, it achieves better system performance only at high SNR.

In \cite{ref17}, a non-iterative minimum mean square error (MMSE)-based channel estimation method is proposed for MIMO-OFDM systems. This method eliminates the need for iterative processing, thereby considerably reducing computational complexity. However, non-iterative methods are prone to error propagation caused by data detection errors. To address this issue, iterative techniques are developed in \cite{ref18}, \cite{ref19}, \cite{ref20}, \cite{ref21}, \cite{ref22}. In \cite{ref18}, an iterative DACE scheme is proposed for both SISO and MIMO orthogonal time frequency space (OTFS) systems. The proposed scheme employs affine-precoded superimposed pilots to enhance channel estimation accuracy. In \cite{ref19}, an iterative channel estimation and data detection algorithm is presented for MIMO-OTFS systems operating in high-mobility environments. This algorithm leverages the sparse nature of the channel in the delay-Doppler domain to iteratively refine both channel estimation and data detection, aiming to enhance spectral efficiency and BER performance. In \cite{ref20}, another iterative DACE scheme is introduced that utilizes soft-decision symbols as pilot signals. Similarly, in \cite{ref21}, an iterative DACE scheme is developed for multicell large antenna systems, employing partially decoded data for channel estimation. Furthermore, in \cite{ref22}, an iterative turbo channel estimation scheme is proposed for OFDM systems, where soft-decision symbols are used as pilot signals in each iteration. However, while these iterative methods address error propagation, they increase computational complexity at the receiver.

\begin{table*}[!t]
\centering
\caption{Related works.}
\label{table:comparison}
\begin{tabular}
{|p{0.08\linewidth}|p{0.255\linewidth}|p{0.255\linewidth}|p{0.31\linewidth}|}
\hline
\textbf{References} & \textbf{System Types} & \textbf{Channel Estimation Methods} & \textbf{Performance Metrics} \\
\hline
\cite{ref14} & MIMO & LMMSE & MSE \\
\hline
\cite{ref15} & SISO and MIMO-OFDM & LS and LMMSE & MSE, BER, and spectral efficiency (SE) \\
\hline
\cite{ref16} & MIMO & LMMSE & MSE and BER \\
\hline
\cite{ref17} & MIMO-OFDM & MMSE & MSE \\
\hline
\cite{ref18} & SISO and MIMO-OTFS & LMMSE & MSE and symbol-error-rate (SER) \\
\hline
\cite{ref19} & MIMO-OTFS & LMMSE & MSE, BER, and SE \\
\hline
\cite{ref20} & MIMO-OFDM & MMSE & MSE and BER \\
\hline
\cite{ref21} & Multicell large antenna systems & LMMSE & BER \\
\hline
\cite{ref22} & OFDM & LS and MMSE & MSE and frame-error-rate (FER) \\
\hline
\cite{ref23} & MIMO & LMMSE & MSE and block-error-rate (BLER) \\
\hline
\cite{ref24} & Multiuser systems & LMMSE & MSE \\
\hline
\cite{ref25} & Massive MIMO & LS & FER \\
\hline
\cite{ref26} & OFDM & LS & MSE and BER \\
\hline
\cite{ref27} & TDS-OFDM & LS and MMSE & MSE and BER \\
\hline
\end{tabular}
\end{table*}

In \cite{ref23}, a semi-data-aided LMMSE channel estimator is proposed for MIMO systems. The fundamental concept of the proposed estimator involves selective exploitation of detected symbol vectors as additional pilot signals. The proposed estimator reduces the computational complexity of iterative data-aided channel estimators. In \cite{ref24}, two semi-blind channel estimation methods are developed to enhance estimation accuracy in multiuser systems by incorporating both pilot and data symbols. These methods utilize Gaussian mixture models which offer computational advantage through parallel processing. In \cite{ref25}, the authors present an iterative LS channel estimation algorithm for a massive MIMO turbo-receiver, which integrates log-likelihood ratios (LLR) from a low-density parity-check (LDPC) decoder with an MMSE estimator to generate soft data symbols. These symbols are further MMSE-weighted and merged with pilot symbols for channel estimation. Similarly, in \cite{ref26}, an iterative DACE scheme is introduced for OFDM systems, employing controlled superposition of training sequences. This approach provides a flexible trade-off between bandwidth efficiency and system MSE and BER performance. Moreover, in \cite{ref27}, a novel low-complexity DACE scheme is proposed for time-domain synchronous (TDS)-OFDM systems. The proposed scheme efficiently improves estimation accuracy without the computational burden of decoding and interleaving and demonstrates robust performance under various transmission conditions, including high mobility and long delay-spread channels. 

In addition to the aforementioned DACE schemes summarized in Table 1, several other DACE schemes have been discussed in the literature. However, to the best of our knowledge, none of them utilizes the peak-power carriers selected at the transmitter of an OFDM system as additional pilot signals, a strategy that can significantly reduce the complexity of the DACE scheme for MIMO-OFDM systems.

\begin{figure*}[tb]
    \centering
    \includegraphics[width=\textwidth]{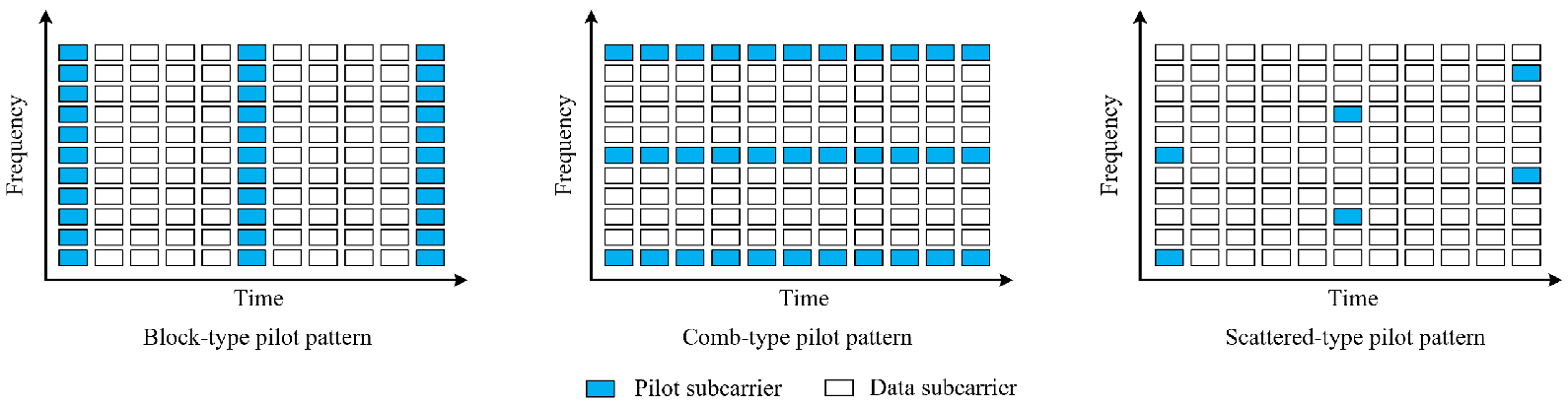}
    \vspace{-0.8em}
    \caption{Pilot patterns.}
    \label{fig1}
    \vspace{-0.8em}
\end{figure*}

\subsection{CONTRIBUTIONS AND NOVELTIES}
In practical wireless communication systems, the received symbols are always different from the transmitted symbols due to severe propagation impairments. Identifying more distorted symbols as reliable can cause severe performance degradation. Simpler detection methods often struggle to detect reliable data symbols at the receiver, whereas more accurate methods can enhance detection performance but at the cost of higher computational complexity. To resolve this trade-off, we propose a novel low-complexity DACE scheme that selects peak-power carriers at the transmitter of an OFDM system. This approach avoids multiple distance calculations at the receiver and significantly reduces system complexity. Peak-power levels occur in an OFDM system when the subcarriers align in phase and constructively interfere with each other. Although high peak-to-average power ratio (PAPR) is generally considered a drawback in OFDM systems, this study uses peak-power carriers as reliable carriers for the DACE scheme. Moreover, we employ the gamma correction companding (GCC) technique to mitigate high PAPR. This technique compresses the signal amplitudes at the transmitter and expands them back at the receiver, thus not affecting the selection of peak-power carriers for the DACE scheme. These peak-power carriers are incorporated with known pilot symbols to update initial channel estimates for both LS and LMMSE channel estimation methods. In this respect, the proposed DACE scheme outperforms traditional channel estimators in terms of system MSE and BER performance. Furthermore, the identified peak-power carriers allow us to extend pilot spacing, thereby reducing the pilot overhead and improving the spectral efficiency of MIMO-OFDM systems.

The main contributions of our work are listed below: 

\begin{itemize}
    
\item [$\bullet$] A novel DACE scheme is proposed that efficiently selects peak-power carriers at the transmitter of an OFDM system. The proposed estimator significantly reduces the computational complexity of traditional receiver-based DACE schemes that selectively exploit detected symbol vectors as additional pilot signals.

\item [$\bullet$] GCC technique is used to reduce high PAPR in OFDM by compressing the signal's amplitude range at the transmitter and expanding it back at the receiver. This process lowers peak values, improving power amplifier efficiency and minimizing distortion during transmission. However, a trade-off is also calculated between the PAPR reduction and performance of channel estimation.

\item [$\bullet$] The proposed DACE scheme is implemented for both SISO and MIMO-OFDM systems. The performance of the proposed channel estimator is evaluated using MSE and BER as the primary metrics over different MIMO configurations, including 1×2, 2×4, and 2×8. 

\item [$\bullet$] Simulation results are obtained for both LS and LMMSE channel estimators for different modulation schemes, including BPSK, 4QAM, and 8PSK. The proposed estimator reduces the pilot overhead by 50$\%$ compared to traditional channel estimators and improves the spectral efficiency of the wireless system.

\end{itemize}

\section{DATA-AIDED CHANNEL ESTIMATION}
As discussed earlier, DACE schemes provide an effective solution for improving channel estimation accuracy by using data symbols as pilot signals to refine initial channel estimates. Among the various factors that directly affect the performance of DACE schemes, the pilot pattern is of primary importance. In this respect, many researchers have proposed joint pilot design and channel estimation schemes for MIMO-OFDM systems, where pilots are inserted in different patterns over OFDM subcarriers. The most commonly used pilot patterns include block-type, comb-type, and scattered-type \cite{ref28}, as shown in Fig. \ref{fig1}. The block-type pilot pattern, in which pilot symbols are inserted on all subcarriers within a specified period, is suitable for slow-fading channels. On the other hand, the comb-type pilot pattern, in which pilot symbols are transmitted on specific subcarriers, is preferable for fast-fading channels. Moreover, the scattered-type pilot pattern, in which pilot symbols are inserted at spaced intervals (on both the time and frequency axes), facilitates time-frequency interpolation techniques. 

In this work, we present a novel DACE scheme using a comb-type pilot pattern to accurately estimate the fast-fading channel with as few pilots as possible. The proposed scheme intelligently selects peak-power carriers at the transmitter of an OFDM system and utilizes them as additional pilot signals to enhance channel estimation accuracy. These peak-power carriers enable us to maintain maximum pilot spacing, resulting in higher spectral efficiency in MIMO-OFDM systems. This is in sheer contrast to the traditional approach of estimating fast-fading channels with a large number of closely spaced pilot symbols. Furthermore, we use the GCC technique to mitigate high PAPR.

\begin{figure*}[tb]
    \centering
    \includegraphics[width=0.9\textwidth]{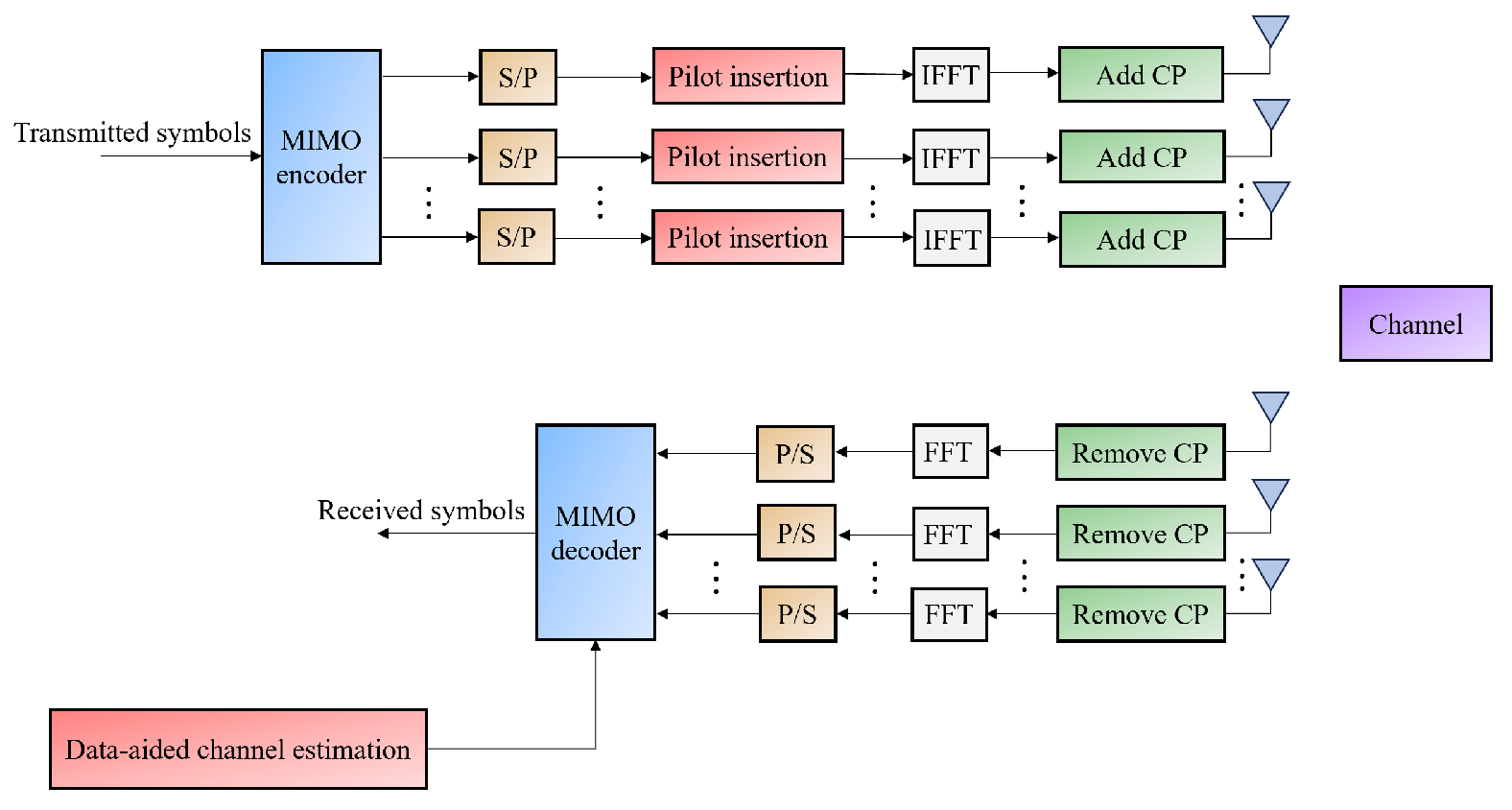} 
    \vspace{-0.8em}
    \caption{MIMO-OFDM system model.}
    \label{fig2}
    \vspace{-0.8em}
\end{figure*}

\subsection{SYSTEM MODEL}
The architecture of a MIMO-OFDM system with parallel inverse fast Fourier transform/fast Fourier transform (IFFT/FFT) operations for the DACE scheme is illustrated in Fig. \ref{fig2}. As an MCM technique, OFDM splits the entire bandwidth into $N$ number of orthogonal subcarriers and converts frequency-selective channel into frequency flat subchannels. The IFFT blocks transform frequency-domain symbols into time-domain signals. Furthermore, each IFFT block uses a cyclic prefix (CP) to prevent inter-symbol interference (ISI). After removing the CP at the receiver, the FFT blocks convert time-domain signals into frequency-domain symbols \cite{ref29}.

The received signal for a MIMO-OFDM system is given by the following equation, where $N_t$ and $N_r$ are the number of transmit and receive antennas, respectively: 
\begin{equation}
\mathbf{y=Hx+z},
\end{equation}
where the vector $\mathbf{x}$ of size $N_t\times1$ represents the transmitted signal on the $\textit{n}^{th}$ subcarrier, $\mathbf{z}$ is the complex AWGN vector with mean zero and covariance matrix $\mathbf{R}_{\mathbf{z}} = {\sigma_{\mathbf{z}}^2} \mathbf{I}_{\textit{N}}$, and matrix $\mathbf{H}$ of size $N_r\times N_t$ represents the channel frequency response (CFR) and is given as:

\begin{equation}
\mathbf{H} = \begin{bmatrix}
 H_{1 1} & \cdots & H_{1 N_t} \\
 \vdots & \ddots & \vdots \\
 H_{N_r 1} & \cdots & H_{N_r N_t}
\end{bmatrix}.
\end{equation}

In matrix-vector form, the received signal is also given as:
\begin{equation}
\mathbf{y=Ch+z},\label{eq3}
\end{equation}
where matrix $\mathbf{C} = \sqrt{N} \mathbf{diag(x)F}$, $\mathbf{diag(x)}$ is an $N \times N$ diagonal matrix hosting the transmitted symbols, $\mathbf{F}$ is the partial FFT matrix of size $N \times L$, and $\mathbf{h}$ represents the $L$-tap channel impulse response (CIR) with tap coefficients $[h_0, h_1, \ldots, h_{L-1}]^\top$.

\subsection{PROPOSED PEAK-POWER-ASSISTED DACE ALGORITHM}
In this subsection, we propose a novel DACE algorithm that selects peak-power carriers at the transmitter of an OFDM system and incorporates them with known pilot symbols to improve channel estimation quality. Unlike traditional approaches, which consider the high PAPR a weakness of an OFDM system, our proposed algorithm turns this weakness into a strength.

In OFDM, data is transmitted on a number of subcarriers spaced at regular intervals. A large number of independently modulated subcarriers are added together to form an OFDM symbol. If these subcarriers are in phase, they can constructively interfere with each other, resulting in high peak-power levels. 

Mathematically, let us consider a simplified scenario where we have \textit{N} subcarriers in an OFDM system. Each subcarrier can be represented as a sinusoidal waveform with a different frequency. Let \textit{x}$_n$(\textit{t}) denote the waveform of the $\textit{n}^{th}$ subcarrier, where \textit{n} = 0, 1, 2, …, \textit{N}-1. The total transmitted signal in the time domain s\textit{}(\textit{t}) can be expressed as the sum of all these individual subcarriers \cite{ref30}, \cite{ref31}:

\begin{equation}
s(t) = \sum_{n=0}^{N-1} x_n(t) = \frac{1}{\sqrt{N}} \sum_{n=0}^{N-1} X_n e^{j 2 \pi f_n t},
\end{equation}
where s(\textit{t}) is the OFDM signal, \textit{N} is the total number of subcarriers, \textit{X}$_n$ is the data on the $\textit{n}^{th}$ subcarrier, and $f_n$ is the frequency of the $\textit{n}^{th}$ subcarrier. The instantaneous power of this signal is: 
$\mid$s(\textit{t})$\mid$ $^2$ = \textit{s}(\textit{t})\textit{s}$^*$(\textit{t}), where \textit{s}$^*$(\textit{t}) is the complex conjugate of \textit{s}(\textit{t}). 

The peak-power is the maximum instantaneous power, and the average power is the expectation of the instantaneous power. Therefore, the PAPR is defined as: max($\mid$\textit{s}(\textit{t})$\mid$$^2$) / E[$\mid$\textit{s}(\textit{t})$\mid$$^2$], where E[$\mid$\textit{s}(\textit{t})$\mid$$^2$] is the expected value of the instantaneous power. Now, if we consider the case where all subcarriers are in phase (i.e., they all have the same phase at any given time), the amplitudes of these subcarriers will add up constructively at certain points in time, leading to a peak in the total power \cite{ref31}, \cite{ref32}.

\begin{figure}[tb]
    \centering
    \includegraphics[width = 1\columnwidth]{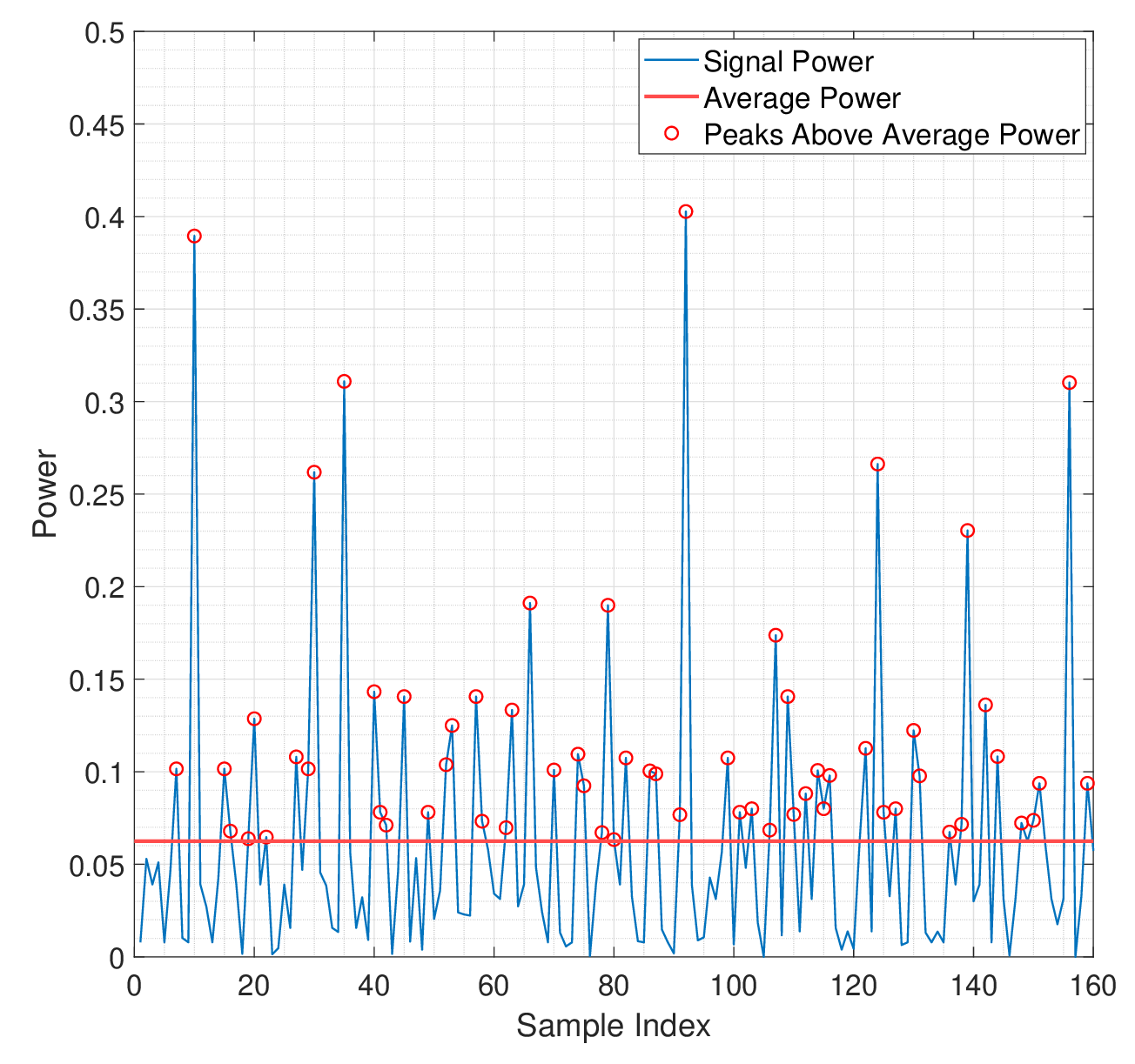} \vspace{-0.8em}
    \caption{Simulating 10 OFDM symbols with \textit{N} = 16 reveals that some subcarriers can have significantly higher power peaks compared to the average power of the signal.}
    \label{fig3}
    \vspace{-0.8em}
\end{figure}

In practical OFDM systems, the subcarriers are not always perfectly in phase due to various factors such as channel fading, Doppler shifts, and timing errors. Therefore, the actual peak-power may be lower than the theoretical peak calculated above. Nevertheless, constructive interference of subcarriers remains a fundamental principle contributing to the peak-power in OFDM systems. Fig. \ref{fig3} illustrates the high PAPR phenomenon in OFDM signals, showing that while the average power is relatively stable, the instantaneous power can have significant peaks.

For an OFDM signal, the autocorrelation function is used to analyze the orthogonality of the subcarriers \cite{ref30}, \cite{ref31}, \cite{ref32}. If the orthogonality of the subcarriers is not maintained (which can be detected using the autocorrelation function), the subcarriers may interfere with each other. This interference can lead to an increase in the peak-power levels, thereby increasing the PAPR. In OFDM systems, high PAPR is considered a drawback due to the increased complexity of analog-to-digital and digital-to-analog conversion processes, as well as the reduced efficiency of power amplifiers. However, we exploit this effect to identify carriers that are less likely to suffer from distortion as they propagate through the wireless channel. Using these less distorted carriers for data transmission could potentially improve system performance. To detect the first \textit{N} carriers with high peak-power in an OFDM system, we use a simple algorithm based on calculating the peak-power of each subcarrier as \textit{P}$_{peak}$ = $\mid$\textit{X}{$_n$}$\mid$$^2$. We then sort the subcarriers by their peak-power values in descending order and select the first \textit{N} subcarriers from the sorted list to identify the required number of carriers.

\subsection{GAMMA CORRECTION COMPANDING TECHNIQUE}
Several techniques have been developed to suppress high PAPR in OFDM signals, including clipping, filtering, error filtering, linear precoding, coding schemes, phase optimization, nonlinear companding, tone reservation, tone injection, constellation shaping, selective mapping, and partial transmit sequence. However, nonlinear companding is particularly promising due to its simplicity and ease of implementation. In this technique, signals with lower amplitudes are increased, while higher amplitudes are compressed. This process normalizes the overall average power of the signals, making it ideal for PAPR reduction in OFDM systems. A-law and $\mu$-law companding functions significantly reduce the dynamic range of the input signal, improving PAPR. In this work, we used GCC for nonlinear companding for its effectiveness and simplicity \cite{ref33}, \cite{ref34}, \cite{ref35}. The mathematical form of the gamma function is a power function, where each input value is raised to a constant power. In OFDM systems, high amplitude occurrences are rare, leading to a low average power and a high PAPR. This can be addressed by reducing peak amplitudes or increasing the average power of the transmitted signal. Gamma correction, covering the dynamic range of all amplitudes, can help. It leaves high-amplitude subcarriers mostly unchanged while amplifying lower-amplitude ones, effectively raising the average power and potentially reducing PAPR. The companding function for discrete-time complex OFDM symbols can be expressed as \cite{ref33}:

\begin{equation}
H_\gamma (x) = A \text{sgn}(x) |x| ^ {1/\gamma}.
\label{eq5}
\end{equation}

In Eq. (\ref{eq5}), \textit{x} denotes the instantaneous value of the OFDM signal, while $\gamma$ represents the companding parameter, and \textit{A} is the amplitude normalization parameter. To negate the effects of gamma correction, GC expanding, an inverse transformation of GCC, is applied at the receiving end. This inverse function is formulated as \cite{ref33}:

\begin{equation}
H^{-1}_\gamma (y) = \text{sgn} (y) \left( \frac{|y| }{A}\right)^ {\gamma}.
\end{equation}

As $\gamma$ approaches 1, the companding curve becomes increasingly linear, necessitating a value of $\gamma$ greater than 1 to introduce nonlinear compression (companding).

\begin{figure}[tb]
    \centering
    \includegraphics[width = 1\columnwidth]{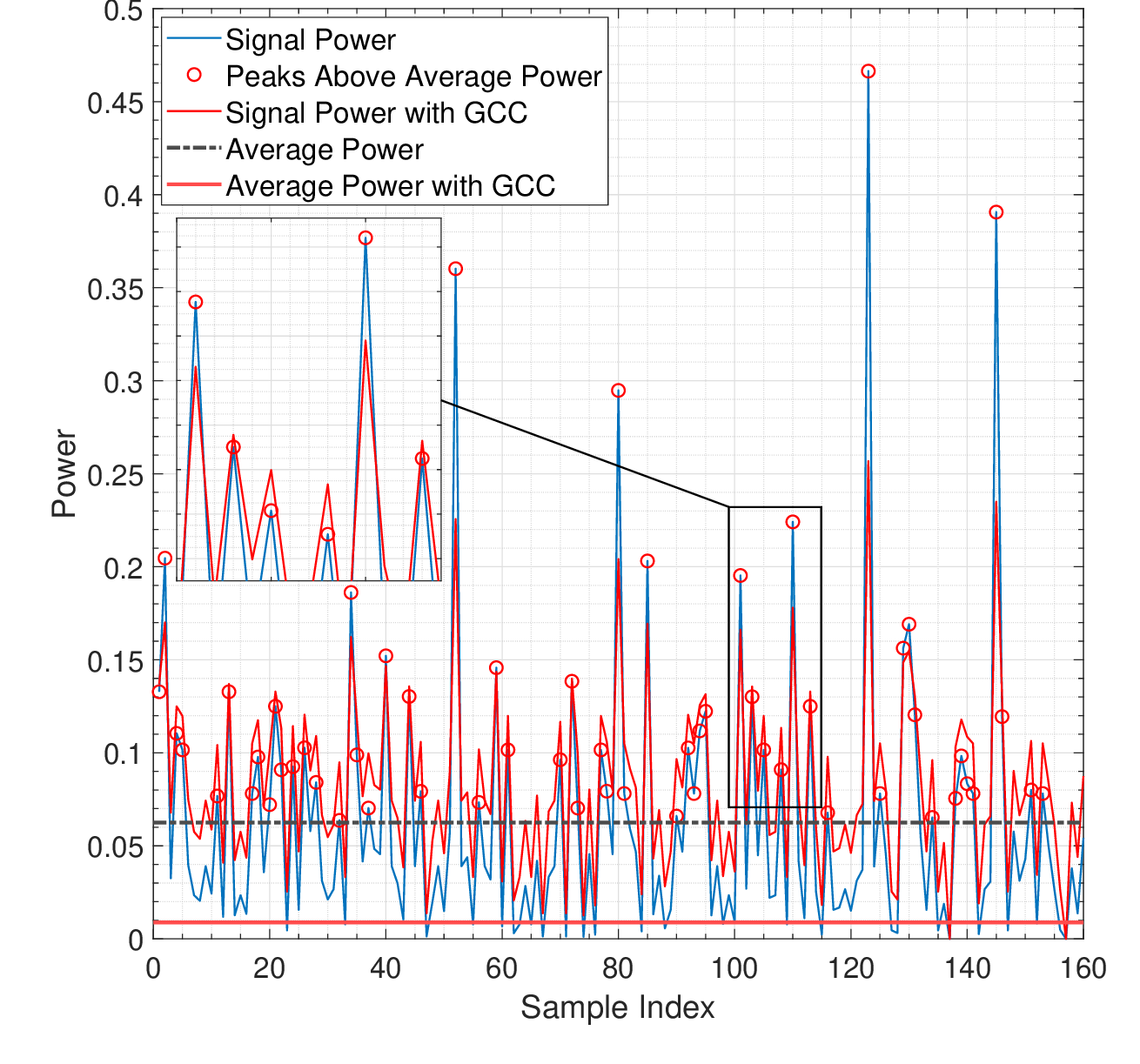} \vspace{-0.8em}
    \caption{Effect of companding on a regular OFDM signal. Larger peaks are compressed and smaller peaks are expanded, reducing the dynamic range and average peak-power.}
    \label{fig4}
    \vspace{-0.8em}
\end{figure}

Fig. \ref{fig4} illustrates the effect of companding on a regular OFDM signal. Here, the larger peaks in the signal are compressed, while smaller peaks are expanded, effectively reducing the dynamic range of the signal. This process leads to a reduction in the average peak-power, improving the signal's efficiency and reducing the likelihood of distortion. The degree of companding can be controlled by the \(\gamma\) factor. As \(\gamma\) increases, higher gain in PAPR can be achieved. However, this also normalizes higher peaks that are used as reliable carriers. Therefore, a trade-off is made in order to achieve an acceptable PAPR reduction as well as distinct peak amplitudes that can be used as reliable carriers.

In this study, we utilize Rapp’s model of the solid-state power amplifier (SSPA), which accurately represents a practical high-power amplifier (HPA). The amplitude/amplitude (AM/AM) and amplitude/phase (AM/PM) characteristics of an SSPA are given by the following equations \cite{ref34}:

\begin{equation}
T(|x(t)|) = \frac{g |x(t)|}{\left[1 + \left(\frac{g|x(t)|}{A_{\text{sat}}}\right)^{2k}\right]^{\frac{1}{2k}}}, \quad \mathit{\Phi}(|x(t)|) = 0 \, \text{(rad)},
\end{equation}
where \( T(\cdot) \) and \( \mathit{\Phi}(\cdot) \) represent the AM/AM and AM/PM conversions, respectively. It is important to note that the SSPA only produces AM/AM conversion and does not introduce any phase distortion. Here, \( |x(t)| \) is the input signal amplitude, \( g \) is the small gain of the SSPA, \( A_{\text{sat}} = \sqrt{P_{\text{sat}}} \) is the output saturation amplitude, \( P_{\text{sat}} \) is the saturated output power, and \( k \) is the knee factor which controls the smoothness of the transition from the linear to the saturation region of the SSPA. 

To avoid nonlinear distortion, the SSPA should operate in the linear region. The output signal of the SSPA can be expressed as:

\begin{equation}
y(t) = T(|x(t)|) e^{j\{\theta(x(t)) + \mathit{\Phi}(x(t))\}}.
\end{equation}
 
Finally, the peak-power carriers are incorporated with known pilot symbols to update the initial channel estimates for both LS and LMMSE channel estimation methods. For $p$ as pilot and $r$ as reliable data indices, Eq. (\ref{eq3}) is updated as: 

\begin{equation}
\mathbf{y}_{rp}=\mathbf{C}_{\textit{rp}}\mathbf{h}+\mathbf{z}_{rp},
\end{equation}
where 
\begin{equation}
\mathbf{C}_{\mathit{rp}} = \sqrt N \mathbf{diag}(\mathbf{x}_{\textit{rp}})\mathbf{F}_{\mathit{rp}}.
\end{equation}
Now, the transmission vector is given as:
\begin{equation}\mathbf{x}_{\textit{rp}}=\begin{bmatrix}
 \mathbf{x}(p) \\
 \mathbf{x}(r)    
\end{bmatrix},\end{equation}
the partial FFT matrix is given as:
\begin{equation}\mathbf{F}_{\textit{rp}}=\begin{bmatrix}
 \mathbf{F}(p) \\
 \mathbf{F}(r)    
\end{bmatrix},\end{equation}
and the observation vector is given as:
\begin{equation}\mathbf{y}_{\textit{rp}}=\begin{bmatrix}
 \mathbf{y}(p) \\
 \mathbf{y}(r)
\end{bmatrix}.\end{equation}

Consequently, the LS and LMMSE channel estimates, utilizing both pilots and peak-power carriers to re-estimate the wireless channel, are given by the following expressions:

\begin{equation}
\hat{\mathbf{h}}_{rp}^{(\text{LS})}=\left(\mathbf{C}_{\textit{rp}}^{\textnormal{H}} \mathbf{C}_{\textit{rp}}\right)^{-1}\mathbf{C}_{\textit{rp}}^{\textnormal{H}}\mathbf{y}_{\textit{rp}},\label{eq14}
\end{equation}

\begin{equation}
\hat{\mathbf{h}}_{rp}^{(\text{L M M S E })}=\left(\mathbf{R}_{\mathbf{h}}^{-1}+\mathbf{C}_{\textit{rp}}^{\textnormal{H}}\mathbf{R}_{\mathbf{z}}^{-1}\mathbf{C}_{\textit{rp}}\right)^{-1}\mathbf{C}_{\textit{rp}}^{\textnormal{H}}\mathbf{R}_{\mathbf{z}}^{-1} \mathbf{y}_{\textit{rp}}.\label{eq15}
\end{equation}

By using peak-power carriers as additional pilot signals, the proposed DACE scheme improves the system MSE and BER performance. Thus, the selection of peak-power carriers in OFDM systems serves as a strategic approach to improve the quality of channel estimation. By identifying subcarriers with the highest peak-power, which inherently have stronger constructive interference, the system prioritizes them as reliable data carriers. This selective approach not only capitalizes on the inherent characteristics of the transmission medium, but also optimizes resource allocation and improves overall system performance. Furthermore, the MIMO-OFDM system effectively utilizes three-dimensional resources, i.e., time, frequency, and space, and offers a high data rate with improved spectral efficiency.

The proposed algorithm is summarized as follows:
\begin{itemize}
    
 \item [$\bullet$] Select the index $r$ of the peak-power carriers at the transmitter of an OFDM system by thresholding i.e., the index vector $r$ is computed as:
    
    for \textit{n} = 0, 1, 2, …, \textit{N}-1,  \textit{n}$\in$ \textit{r}, \textit{iff} \textit{P}$_{\text{peak}}(n) > \text{threshold}$.

  \item [$\bullet$] Employ GCC technique to reduce high PAPR in OFDM by compressing the signal’s amplitude range at the transmitter and expanding it back at the receiver.

  \item [$\bullet$] Get initial channel estimates $\hat{\mathbf{h}}$ using LS and LMMSE channel estimation methods.
        
  \item [$\bullet$] Incorporate the peak-power carriers with known pilot symbols and obtain updated channel estimates $\hat{\mathbf{h}}_{rp}^{(\text{LS})}$ and $\hat{\mathbf{h}}_{rp}^{(\text{L M M S E})}$, using (\ref{eq14}) and (\ref{eq15}), respectively.

  \item [$\bullet$] Analyze the system MSE and BER performance for the transmitter-based DACE scheme and compare it with the receiver-based DACE scheme.
    
  \item [$\bullet$] Ascertain the improvement in the spectral efficiency of the wireless system using both transmitter-based and receiver-based DACE schemes.
  
\end{itemize}

\subsection{COMPLEXITY ANALYSIS}
The computational complexity of both the transmitter-based and receiver-based DACE schemes is influenced by several factors, including the number of OFDM subcarriers $N$, matrix multiplication and inversion operations, the number of Monte Carlo trials, and the range of SNR values. However, we are interested in a comparative analysis of the computational complexity involved in selecting reliable data carriers for both the transmitter-based and receiver-based DACE schemes.

For the transmitter-based DACE scheme, the complexity is represented by $\mathcal{O}(N \log N)$, which is primarily due to the selection of peak-power carriers at the transmitter of an OFDM system. In contrast, the receiver-based DACE scheme, proposed in \cite{ref15}, which selects reliable data carriers at the receiver of an OFDM system, features a complexity of $\mathcal{O}(N \cdot M + N \log N)$, where $M$ represents the size of the constellation set. 

The comparative analysis of the computational complexity is given by the following expression:

\begin{equation}
\text{Complexity Ratio} = \frac{\mathcal{O}(N \log N)}{\mathcal{O}(N \cdot M + N \log N)} \approx \frac{1}{M}. 
\end{equation}

This ratio indicates that the computational load of the receiver-based DACE scheme is higher by approximately a factor of $M$ due to the extensive reliability calculations required for each data carrier. Therefore, while the receiver-based approach may offer higher accuracy and robustness, it demands significantly more computational resources, a consideration that is crucial in the design and implementation of DACE schemes in OFDM systems.

\begin{table}[!t]
\centering
\caption{\textbf{Simulation parameters.}}
\label{my-label}
\begin{tabular}{l @{\hspace{1.75cm}} l} 
\toprule
\addlinespace[-0.65ex]
\toprule
\textbf{Parameter} & \textbf{Specification} \\ 
\midrule
Number of OFDM subcarriers & $N = 256$ \\
Number of pilot subcarriers & $N_p = 16$ \\
Number of data subcarriers & $N_d = 240$ \\
Channel model & Rayleigh fading \\
Number of channel taps & $L = 16$ \\
Pilot pattern & Comb-type \\
Channel estimation & DACE scheme \\
Channel estimation methods & LS and LMMSE \\
Modulation schemes & BPSK, 4QAM, and 8PSK \\
MIMO configurations & $1 \times 2$, $2 \times 4$, and $2 \times 8$ \\
Target MSE & 0.02 \\ 
\bottomrule
\addlinespace[-1.15ex]
\bottomrule
\end{tabular}
\end{table}

\begin{figure}[tb]
    \centering
    \includegraphics[width = 1.08\columnwidth]{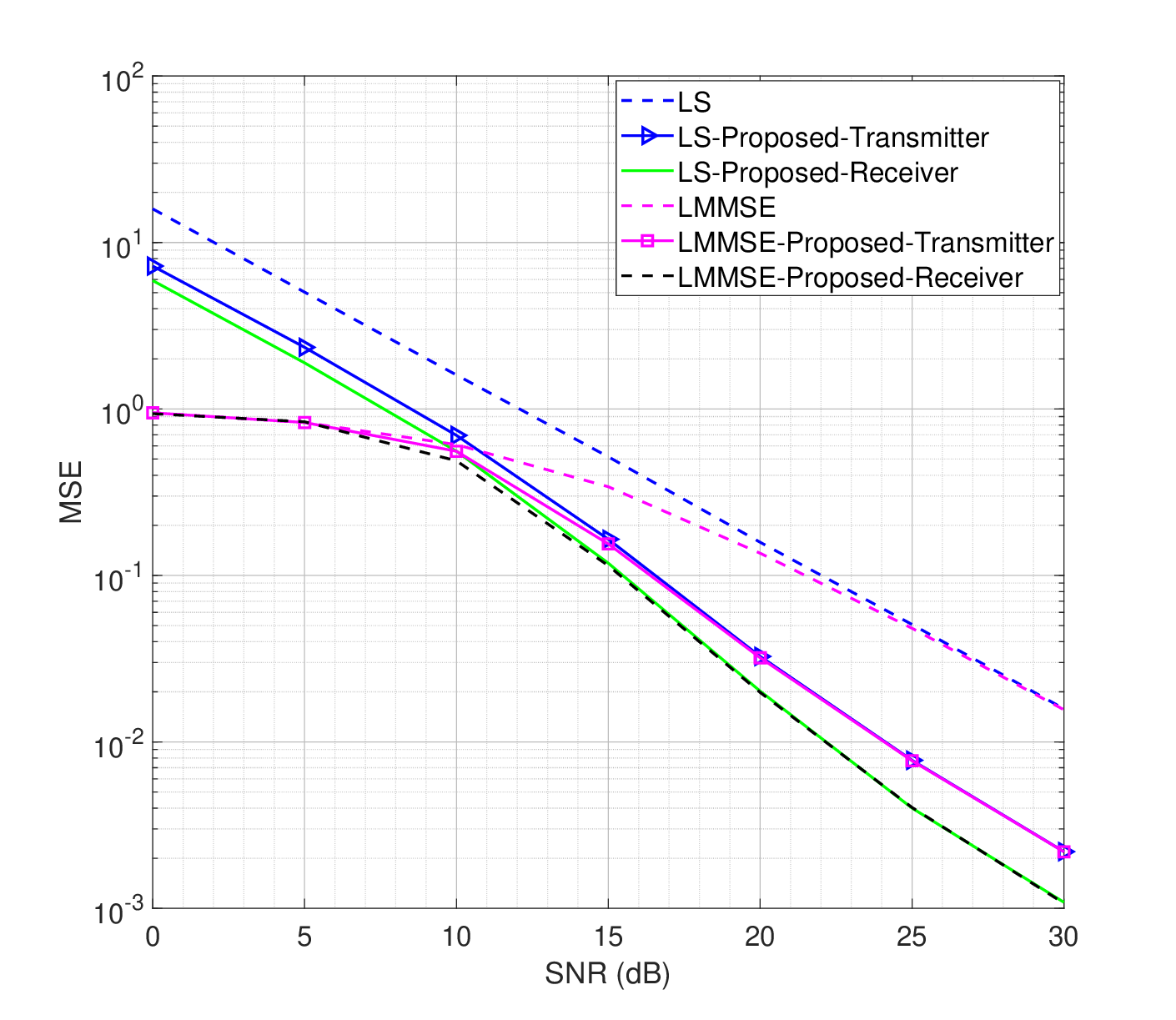} \vspace{-0.8em}
    \caption{MSE versus SNR for BPSK, $N_t = 1$, $N_r = 1$.}
    \label{fig5}
    \vspace{-0.8em}
\end{figure}

\begin{figure}[tb]
    \centering
    \includegraphics[width = 1.08\columnwidth]{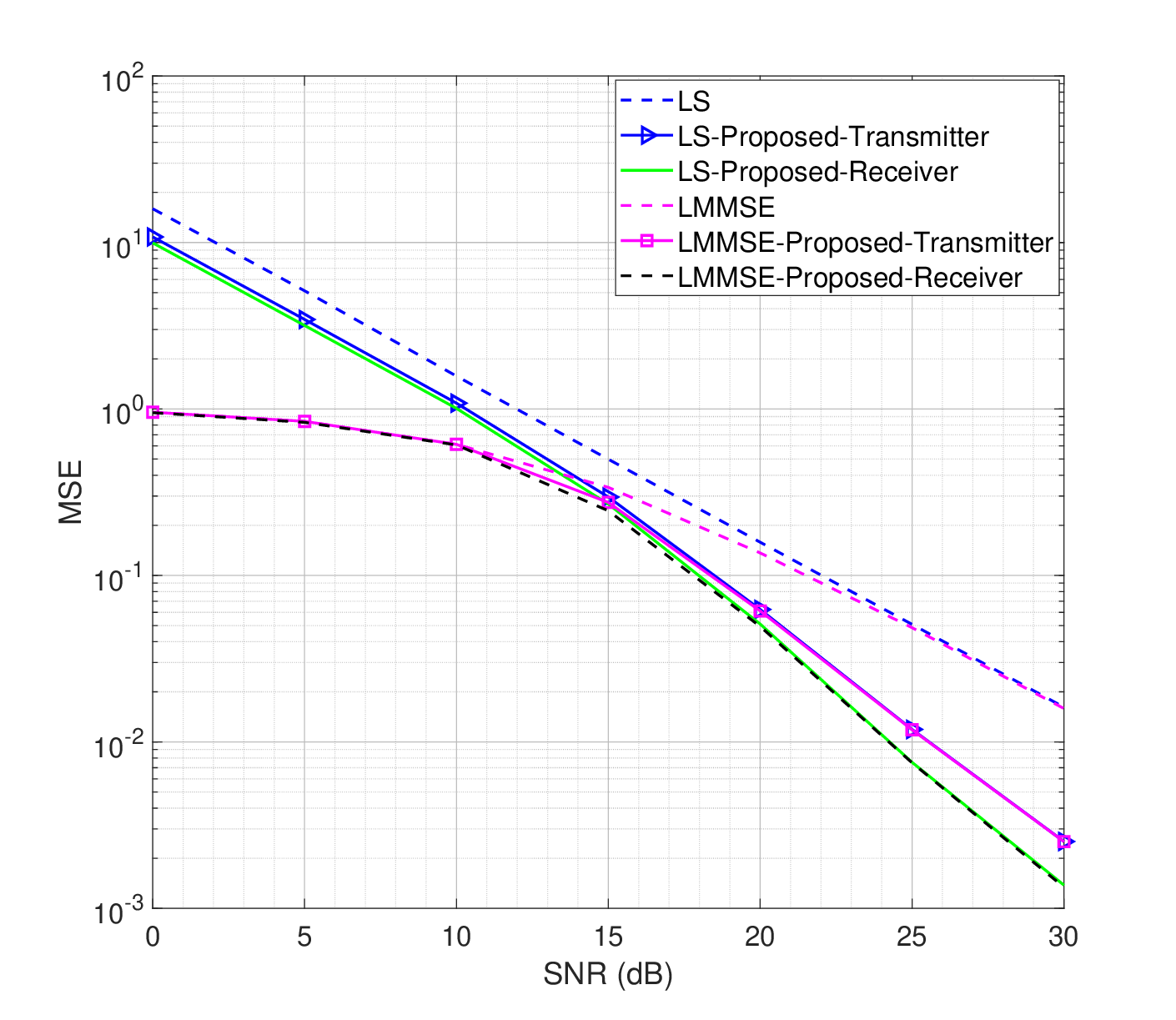} \vspace{-0.8em}
    \caption{MSE versus SNR for 4QAM, $N_t = 1$, $N_r = 1$.}
    \label{fig6}
    \vspace{-0.8em}
\end{figure}

\begin{figure}[tb]
    \centering
    \includegraphics[width = 1.08\columnwidth]{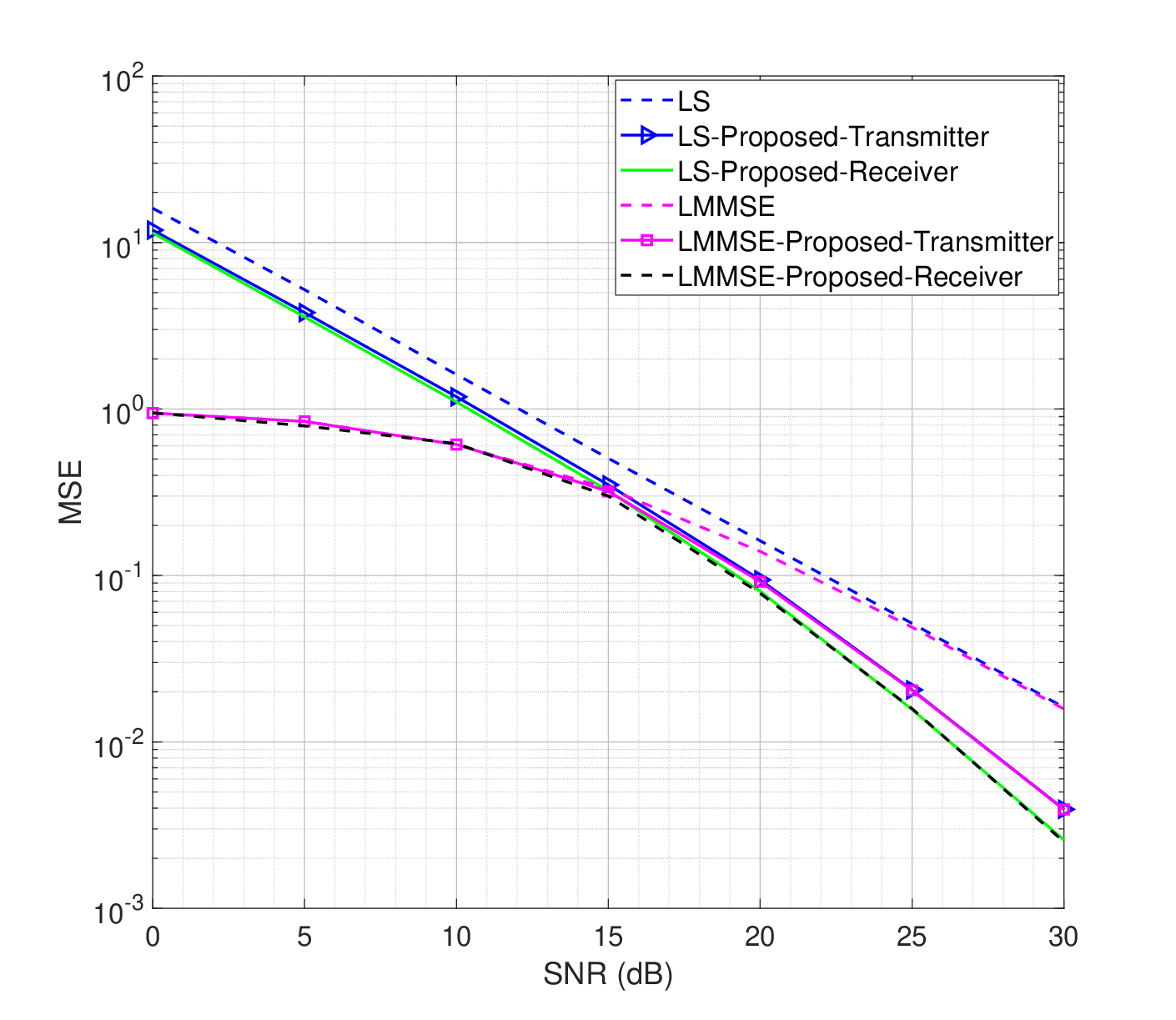} \vspace{-0.8em}
    \caption{MSE versus SNR for 8PSK, $N_t = 1$, $N_r = 1$.}
    \label{fig7}
    \vspace{-0.8em}
\end{figure}

\begin{figure}[tb]
    \centering
    \includegraphics[width = 1.08\columnwidth]{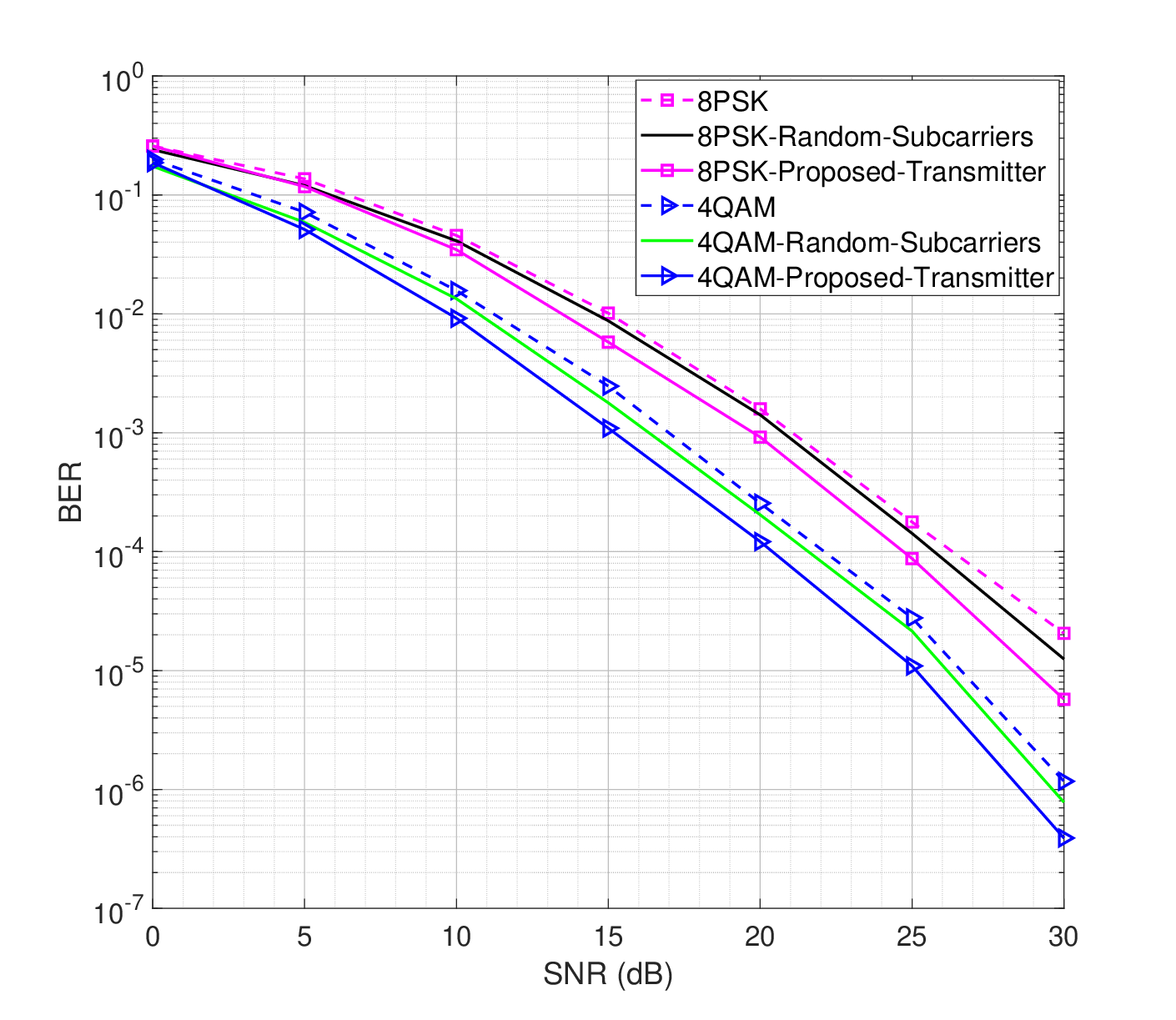} \vspace{-0.8em}
    \caption{BER versus SNR for LS/LMMSE, $N_t = 1$, $N_r = 2$.}
    \label{fig8}
    \vspace{-0.8em}
\end{figure}

\begin{figure}[tb]
    \centering
    \includegraphics[width = 1.08\columnwidth]{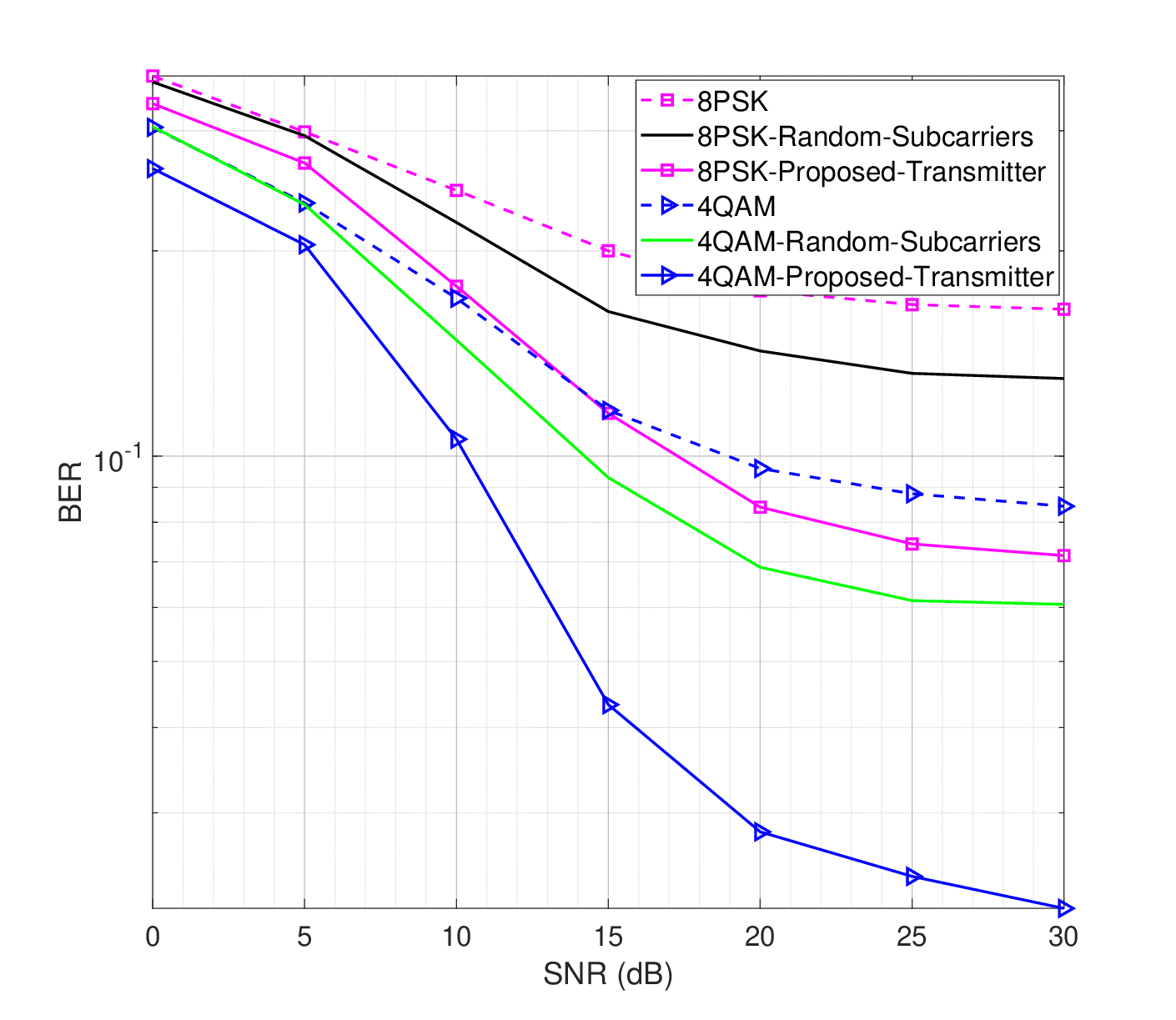} \vspace{-0.8em}
    \caption{BER versus SNR for LS/LMMSE, $N_t = 2$, $N_r = 4$.}
    \label{fig9}
    \vspace{-0.8em}
\end{figure}

\begin{figure}[tb]
    \centering
    \includegraphics[width = 1.08\columnwidth]{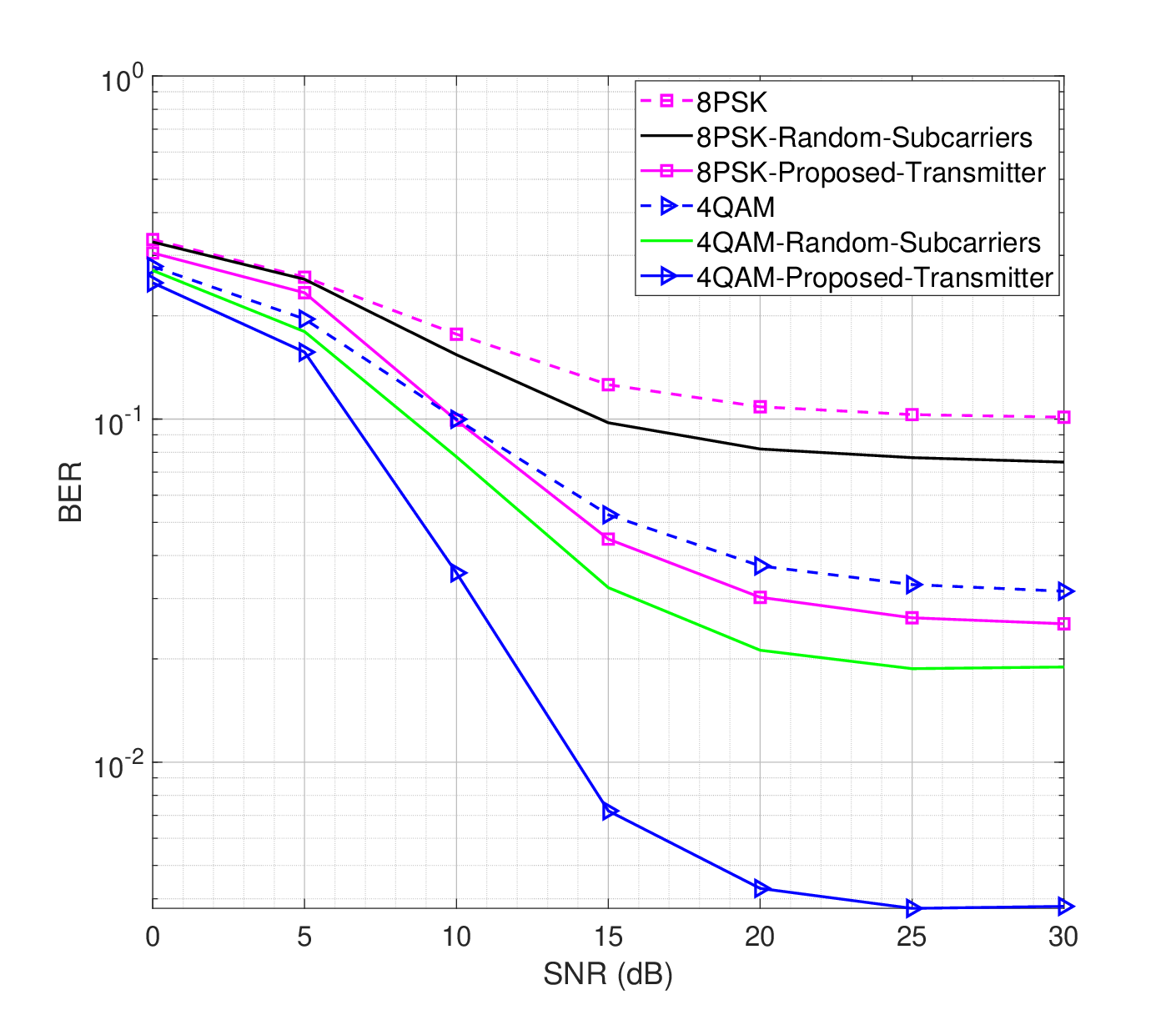} \vspace{-0.8em}
    \caption{BER versus SNR for LS/LMMSE, $N_t = 2$, $N_r = 8$.}
    \label{fig10}
    \vspace{-0.8em}
\end{figure}

\section{RESULTS AND DISCUSSION}
In this section, we present the simulated results obtained using the parameters listed in Table 2. We compare the performance of the proposed transmitter-based DACE scheme with traditional LS and LMMSE channel estimators, the receiver-based DACE scheme developed in \cite{ref15}, and channel estimators using random subcarriers as additional pilot signals. Using an identical number of reliable data carriers for both the transmitter-based and receiver-based DACE schemes, the MSE curves are plotted against SNR to analyze system performance. The MSE performance for the SISO system using both LS and LMMSE channel estimators for BPSK, 4QAM, and 8PSK modulation schemes is shown in Figs. \ref{fig5}, \ref{fig6}, and \ref{fig7}, respectively. The MSE curves for both the proposed transmitter-based and receiver-based schemes illustrate the performance of channel estimators using both pilots and reliable data carriers. It is observed that reliable data carriers significantly improve system performance for both schemes. However, the transmitter-based scheme achieves equivalent MSE performance compared to the receiver-based scheme, suggesting that both schemes extract nearly the same reliable data carriers. Furthermore, lower-order constellations achieve better performance than higher-order constellations because they typically have fewer, more widely spaced symbols. This greater spacing between constellation points enhances their robustness to both noise and interference. Therefore, BPSK, with its maximum phase separation, achieves better performance than 4QAM, and 4QAM typically outperforms 8PSK under similar conditions.

The BER is now calculated as the number of errors found in the transmitted symbols divided by the total number of transmitted symbols. Using the proposed transmitter-based DACE scheme, the BER curves are plotted for 4QAM and 8PSK modulation schemes for 1×2, 2×4, and 2×8 MIMO configurations, as shown in Figs. \ref{fig8}, \ref{fig9}, and \ref{fig10}, respectively. The BER performance of the proposed scheme is compared to both traditional channel estimators and channel estimators using random subcarriers as additional pilot signals. In this regard, Fig. \ref{fig8} demonstrates that the proposed scheme outperforms both types of channel estimators. Similar behaviour is observed for the 2×4 and 2×8 MIMO configurations, as illustrated in Figs. \ref{fig9} and \ref{fig10}, respectively. However, the 2x8 MIMO configuration offers superior performance compared to the 2x4 MIMO configuration, primarily due to the diversity gain. Increasing the number of receive antennas enhances the robustness of signal detection and decoding. This diversity gain lowers the error rate by providing multiple observations of the same transmitted signal, allowing for improved mitigation of noise and channel fading effects. Moreover, the MSE and BER performance appears to be better at high SNR as the data equalization/channel estimation has less distortion at high SNR and vice versa.

\begin{figure}[tb]
    \centering
    \includegraphics[width = 1.08\columnwidth]{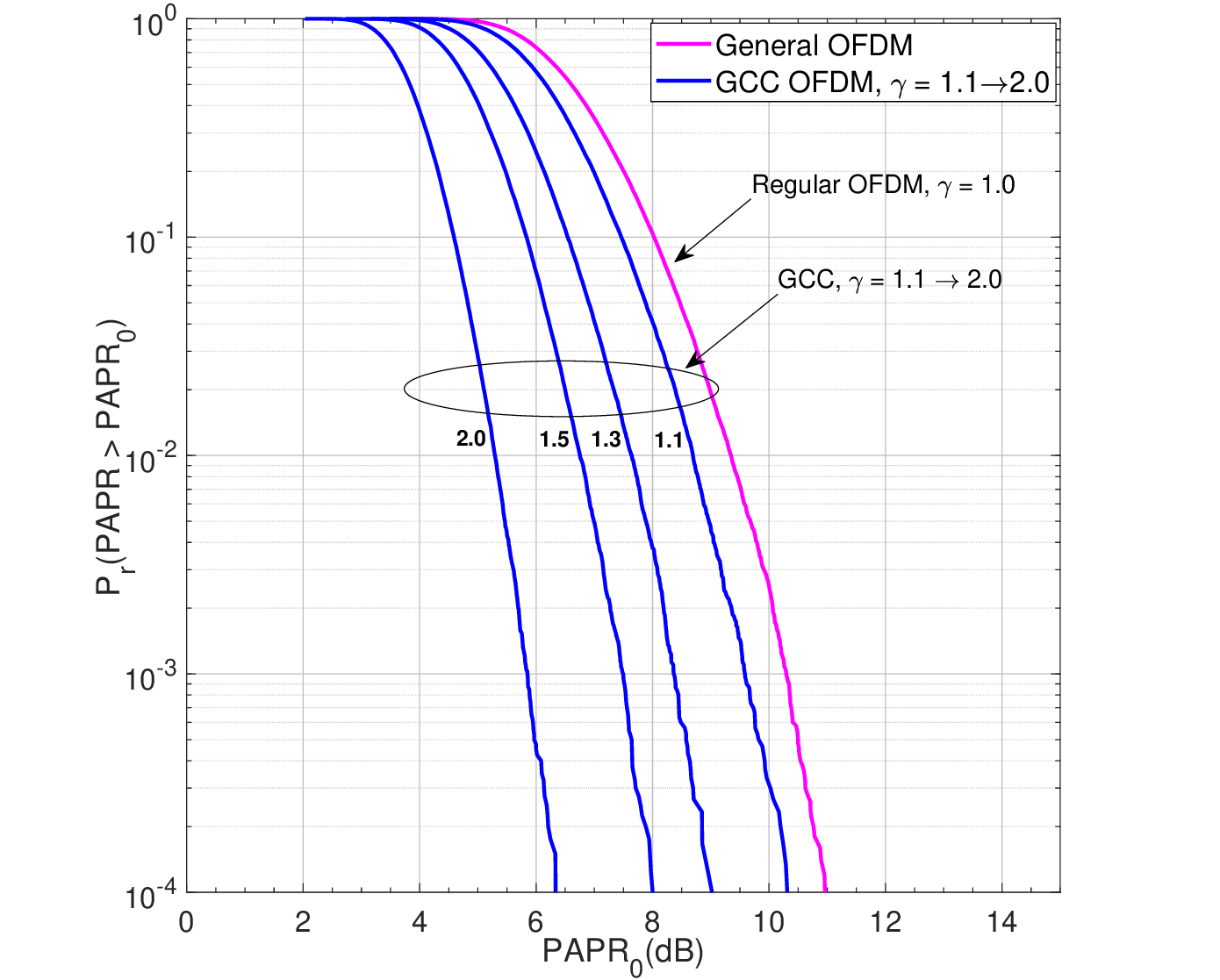} \vspace{0.8em}
    \caption{CCDF curves for General OFDM and GCC-OFDM.}
    \label{fig11}
    \vspace{-0.8em}
\end{figure}

In general, companding techniques in OFDM systems introduce nonlinear distortions, which can adversely affect BER performance. The degree of nonlinear distortion typically increases with higher PAPR gains, and this distortion can be mitigated by adjusting the gamma parameter \cite{ref35}. However, integrating companding techniques with OFDM—by compressing the dynamic range of the signal before transmission—can enhance the signal's average power relative to the noise (refer to Fig. \ref{fig4}) and improve transmission robustness, especially in noisy environments. The increase in signal power resulting from nonlinear distortion is a consequence of the PAPR reduction technique. We assessed the performance of PAPR reduction using the complementary cumulative distribution function (CCDF) of PAPR, referred to as 'PAPR CCDF'. This metric quantifies the probability that the PAPR of an OFDM sequence exceeds a specified threshold level, \( \text{PAPR}_0 \). Fig. \ref{fig11} illustrates the PAPR CCDF for an OFDM system without PAPR reduction, referred to as 'General OFDM', and for OFDM utilizing GCC companding, denoted as 'GCC OFDM', with 8PSK modulation and \( N = 256 \) FFT. Nonlinear distortions introduced by companding can negatively impact system error performance, with these distortions increasing as the parameter \( \gamma \) rises. A balance between error performance and PAPR reduction is achievable when \( \gamma \) values are selected within the range of 1.1 to 2.0. Although higher \( \gamma \) values increase PAPR gain, they also significantly degrade error performance. In the simulations presented, \( \gamma = 2.0 \) was chosen, demonstrating a substantial improvement in PAPR reduction (approximately 4.5 dB) while enhancing BER performance.

\begin{figure}[tb]
    \centering
    \includegraphics[width = 1.08\columnwidth]{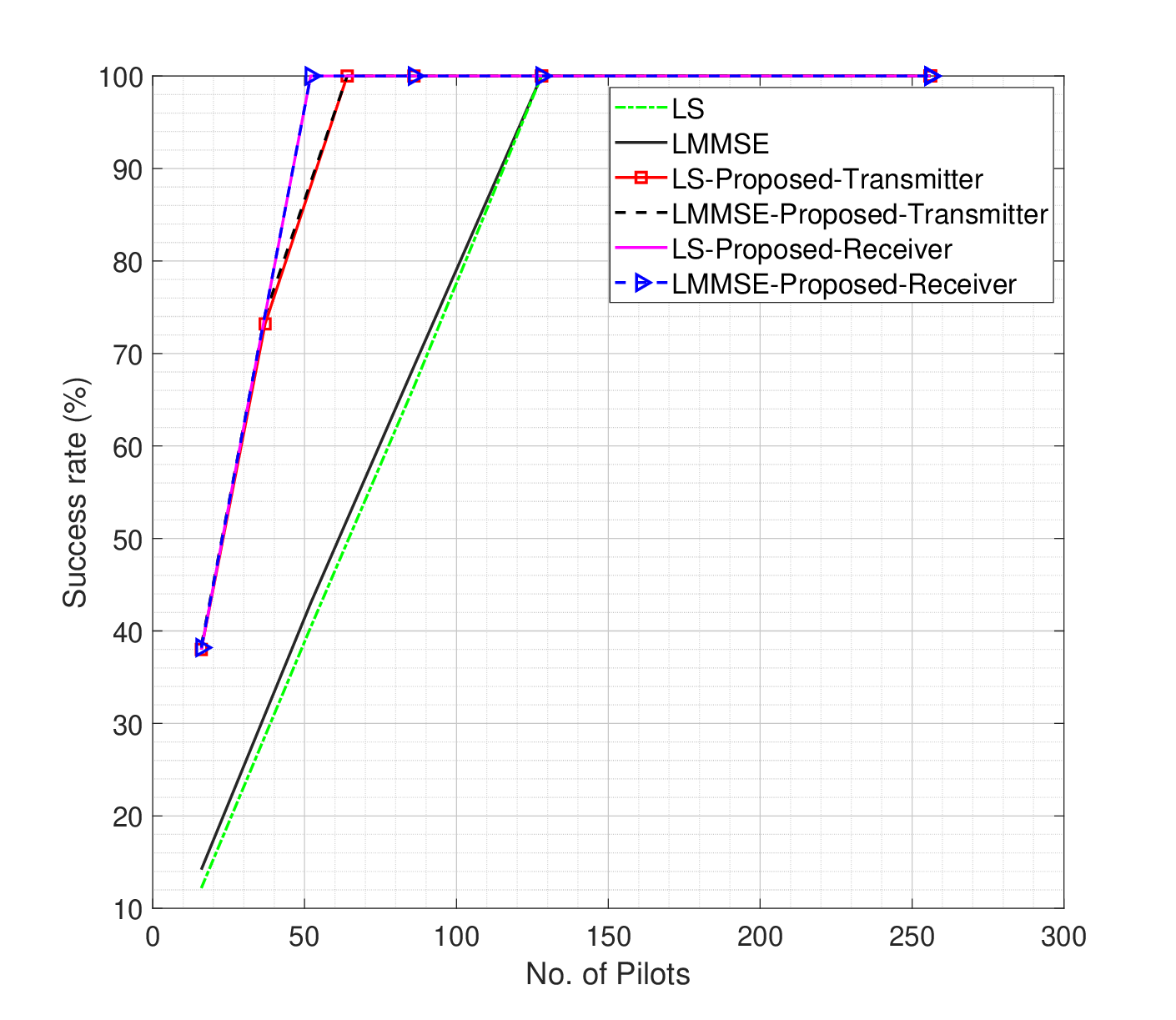} \vspace{-0.8em}
    \caption{Success rate versus number of pilots.}
    \label{fig12}
    \vspace{-0.8em}
\end{figure}

In Fig. \ref{fig12}, the success rate is plotted against the number of pilots to determine the improvement in the spectral efficiency of the system using both transmitter-based and receiver-based DACE schemes. The success rate is defined as the number of symbols with MSE values less than the target MSE divided by the total number of transmitted symbols. It is expressed in percent for the 4QAM constellation at a fixed SNR of 20 dB. It is observed that traditional LS and LMMSE channel estimators achieve a 100$\%$ success rate with 128 number of pilots. However, the proposed LS and LMMSE channel estimators, using the receiver-based DACE scheme, achieve a 100$\%$ success rate with only 52 number of pilots. On the other hand, the proposed LS and LMMSE channel estimators, using the transmitter-based DACE scheme, achieve a 100$\%$ success rate with 64 number of pilots. Therefore, the receiver-based DACE scheme reduces the pilot overhead by approximately 60$\%$, and the transmitter-based DACE scheme reduces it by 50$\%$, compared to traditional channel estimators. Consequently, the proposed transmitter-based DACE scheme achieves comparable performance with significantly lower computational complexity and enhances the spectral efficiency of the wireless system.

\section{CONCLUSION}
In this paper, for the first time, we propose a joint PAPR reduction and low-complexity peak-power-assisted DACE scheme for both SISO and MIMO-OFDM wireless systems. The proposed scheme judiciously selects peak-power carriers at the transmitter of an OFDM system and uses them as additional pilot signals to accurately estimate the fast-fading channel. The GCC technique, which compresses the signal amplitudes at the transmitter and expands them back at the receiver, ensures sufficient margin for the DACE scheme. This way, the proposed DACE scheme not only improves channel estimation accuracy but also enhances the spectral efficiency of MIMO-OFDM systems. Simulation results demonstrate that the proposed scheme outperforms traditional LS and LMMSE channel estimators in terms of system MSE and BER performance. It achieves equivalent MSE and BER performance compared to conventional receiver-based DACE schemes but with significantly lower computational complexity. Furthermore, it efficiently utilizes bandwidth and reduces the pilot overhead by 50$\%$ compared to traditional channel estimators, thereby enhancing the spectral efficiency of the wireless system. 

\bstctlcite{IEEEexample:BSTcontrol}
\bibliographystyle{IEEEtran}

\bibliography{IEEEabrv, paper.bib}

\vskip -2\baselineskip plus -1fil
\balance

\vfill

\end{document}